\documentclass[aps,prl,superscriptaddress,twocolumn,notitlepage]{revtex4-1}
\usepackage[latin1]{inputenc}
\usepackage{amsmath}
\usepackage{amsfonts}
\usepackage{amssymb}
\usepackage{graphicx}
\usepackage{epstopdf}
\graphicspath{ {images/} }
\usepackage[titletoc]{appendix}
\usepackage{mathrsfs}
\usepackage{subfigure}
\usepackage{hyperref}

\usepackage{color}

\begin{document}

	\title{Minimal model of charge and pairing density waves in X-ray scattering experiments }
	
	\author{David Dentelski}
	\affiliation{Department of Physics, Bar-Ilan University, 52900, Ramat Gan Israel}
	\affiliation{Center for Quantum Entanglement Science and Technology, Bar-Ilan University, 52900, Ramat Gan Israel}

	\author{Emanuele G.~Dalla Torre}
	\affiliation{Department of Physics, Bar-Ilan University, 52900, Ramat Gan Israel}
	\affiliation{Center for Quantum Entanglement Science and Technology, Bar-Ilan University, 52900, Ramat Gan Israel}

	\begin{abstract}
		Competing density waves play an important role in the mystery of high-temperature cuprates superconductors. In spite of the large amount of experimental evidence, the fundamental question of whether these modulations represent charge or pairing density waves (CDWs or PDWs) is still debated. Here we present a method to answer this question using both momentum and energy-resolved resonant X-ray scattering maps. Starting from a minimal model of density waves in superconductors, we identify distinctive signatures of incipient CDWs and PDWs. The generality of our approach is confirmed by a self-consistent solution of an extended Hubbard model with attractive interaction. By considering the available experimental data, we claim that the spatial modulations in cuprates have a predominant PDW character.  Our work paves the way for using X-ray to identify competing and intertwined orders in superconducting materials. 
	\end{abstract}
	
	\maketitle
	
	\newpage

	 {\it Introduction --} Strongly correlated materials  often exhibit competing phases with distinct charge and spin orders. A famous example is copper-oxide high-temperature superconductors, or cuprates, whose rich phase diagram poses many theoretical challenges.
	Since the discovery of unidirectional  spin density waves in LSCO \cite{tranquada1995evidence}, it has become increasingly accepted that in cuprates superconductivity  is intertwined with other orders \cite{doiron2007quantum,wu2011magnetic,fradkin2012high,leboeuf2013thermodynamic,wu2013emergence}. In particular, in 2002, scanning tunneling experiments found incommensurate density waves on the surface of BSCCO \cite{hoffman2002four, howald2003periodic, vershinin2004local, hanaguri2004checkerboard}. Ten years later, resonant X-ray scattering experiments detected a similar incommensurate order in the bulk of YBCO \cite{ghiringhelli2012long}. The same order was later found in a large number of cuprates, demonstrating that this effect is ubiquitous \cite{chang2012direct,
		torchinsky2013fluctuating,blackburn2013x,comin2014charge,da2014ubiquitous,le2014inelastic,hashimoto2014direct,tabis2014charge,huecker2014competing,achkar2014impact,gerber2015three,hamidian2015magnetic,peng2016direct, chaix2017dispersive, peng2018re,  jang2017superconductivity,da2018coupling,bluschke2019adiabatic, kang2019evolution}.

	In spite of the large number of experimental studies, the physical interpretation of these periodic modulations is still debated. A common approach, also based on earlier theoretical predictions \cite{castellani1995singular, castellani1997charge}, claims that these modulations are due to a charge density wave (CDW) order that competes with superconductivity.
	While this approach is widely accepted in the literature, it is inconsistent with some experimental details. 
	In  particular, angle-resolved photoemission spectroscopy (ARPES) shows that these density waves are associated with a spectral gap that closes from below the Fermi energy \cite{he2011single},  while CDWs' gaps are expected to close from above. 
	Accordingly, it was argued that the competing order is intimately related to superconductivity \cite{loret2019intimate} and thus interpreted as a pair density wave (PDW) \cite{chen2004pair, lee2014amperean}, or a CDW/PDW mixed order \cite{pepin2014pseudogap, freire2015renormalization, wang2015interplay, wang2015coexistence}. This claim is also supported by recent scanning measurements in the halos of magnetic vortices \cite{edkins2019magnetic} and with superconducting tips \cite{hamidian2016detection,du2020imaging} (see Ref.~\cite{agterberg2019physics} for a recent review).
	
	Here we address the question of how to distinguish between CDW and PDW modulations in available X-ray scattering experiments. Our approach departs from earlier studies that focused on the normal state of cuprates \cite{dalla2016friedel,arpaia2019dynamical} and included the effects of strong antiferromagnetic fluctuations \cite{sachdev2013bond} and Fermi arcs with hot spots \cite{efetov2013pseudogap,allais2014density,pepin2014pseudogap,wang2015interplay, freire2015renormalization,wang2015interplay,wang2015coexistence}.
	 Instead, we base our analysis on the well-established description of the superconducting state of cuprates in terms of a Bardeen-Cooper-Schrieffer (BCS) Hamiltonian with a $d$-wave gap. In this phase, superconductivity suppresses competing orders,  justifying a weak-coupling approach where the density waves are induced by weak pinning centers \cite{caplan2015long, caplan2017dimensional}. By considering a minimal model of isotropic scatterers, we develop a method to distinguish between incipient CDW and PDW fluctuations, which become long-ranged at high magnetic fields.
	The validity of this approach is confirmed by the solution of an extended Hubbard model with attractive interactions in the presence of local impurities, which enables us to study the interplay between CDWs and PDWs.

	\begin{figure*} [t] 
		\centering
		\begin{tabular}{c c c}
			(a) NCCO  & (b) Hg1201 & (c) BSCCO\\
			\includegraphics[width=0.31\textwidth]{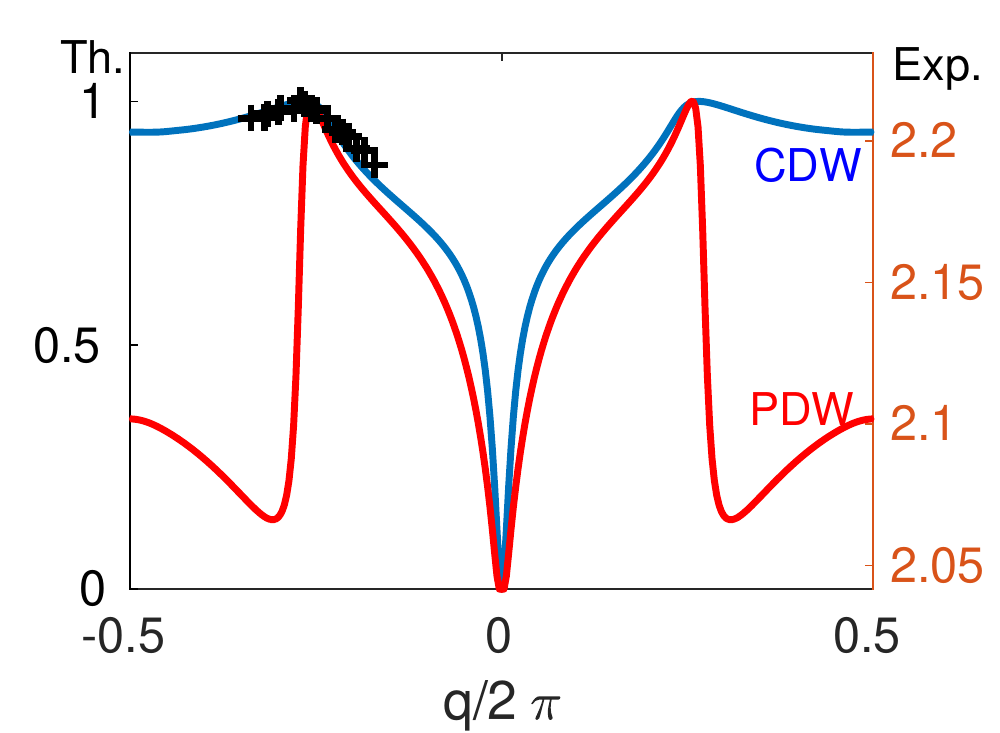}&
			\includegraphics[width=0.31\textwidth]{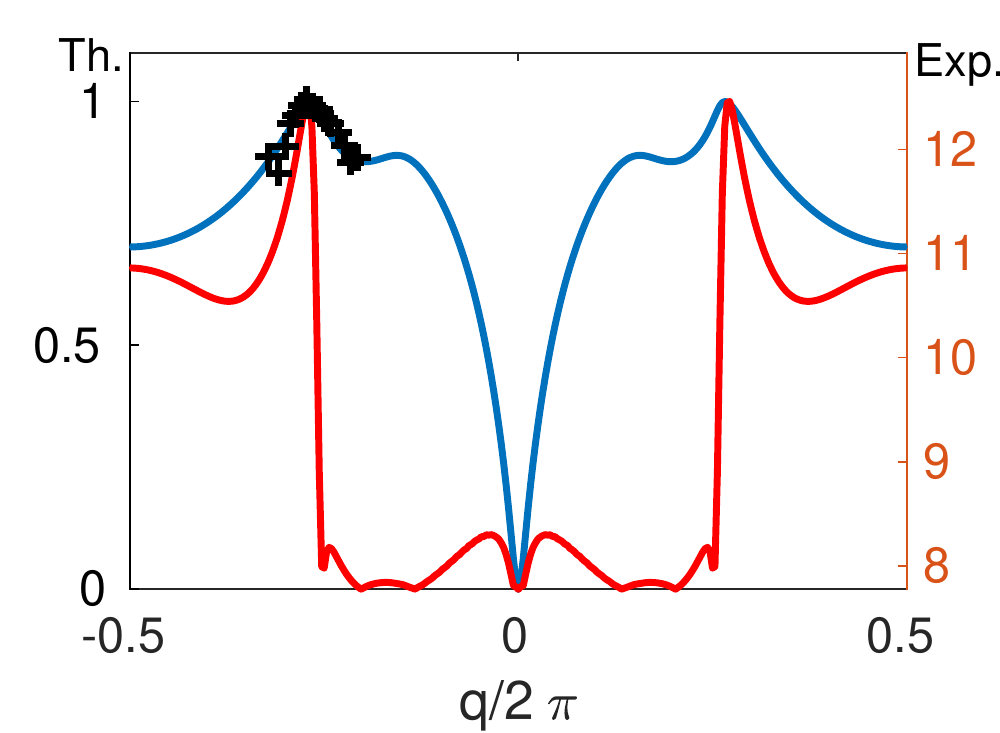}&
			\includegraphics[width=0.31\textwidth]{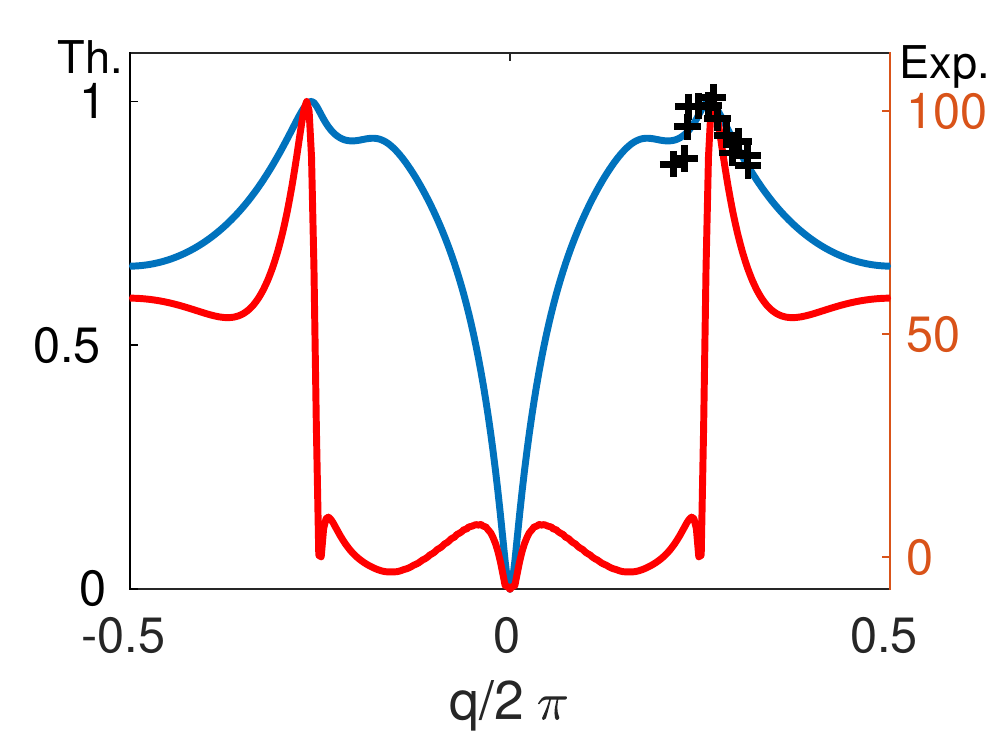}
		\end{tabular}
		\caption {Momentum dependence of the theoretical (Th. -- continuous lines) and experimental (Exp. -- crosses) elastic X-ray scattering signal in the $(q,0)$ direction. The theoretical curves of incipient CDWs (PDWs) are shown in blue (red) and represent Eq.~(\ref{eq:rhoq2}) [Eq.~(\ref{eq:rhoq3})] with $\Omega=0$, $\Delta_0/t=0.1 \ (\Delta_0/t=0.3)$ and (a) $t'/t=-0.22  \ (-0.4)$, (b) $ t'/t=-0.7 \ (-0.7)$, (c) $t'/t=-0.7 \ (-0.7)$. Each theoretical curve is normalized by its maximal value. The experimental data is reproduced from (a) electron doped NCCO, $x=-0.14$ \cite{jang2017superconductivity}; (b) underdoped Hg1201, $x=0.09$ \cite{tabis2014charge}; (c) underdoped BSCCO, $x=0.12$ \cite{comin2014charge}}.
		\label{Fig1}
	\end{figure*} 
	
	 {\it Weak coupling approach --} Resonant X-ray scattering experiments probe density fluctuations at a fixed wavevector $\textbf{q}$ and frequency $\Omega$. In our weak-coupling approach we assume that  incipient CDWs and PDWs can be modeled by a homogeneous state perturbed by a local pinning center (impurity). Under this approximation, the intensity of the X-ray signal is given by density response to the impurity, $\chi(\textbf{q},\Omega)$. In a BCS superconductor, one has (see, for example, Ref.\cite{altland2010condensed})
	\begin{align} \label{eq:rhoq}
	\chi(\textbf{q},\Omega) = \int d\omega \int d^{d}k~\rm Tr \left[G_{0}(\textbf{k},\omega) V_{\bf k} G_{0}(\textbf{k}+\textbf{q}, \omega+\Omega)\sigma^z\right].
	\end{align}
	Here $\rm V$ models a static (time independent) and local ($q$ independent) impurity, $\sigma^z$ is a Pauli matrix, and  $\rm G_0$ is the Green's function
	\begin{align} 
	\rm G^{-1}_{0}({\bf k},\omega)=
	\begin{pmatrix} - \omega +\varepsilon_{\textbf{k}}-\mu& \Delta_{\textbf{k}} \\  \Delta^{\star}_{\textbf{k}} &-\omega-\varepsilon_{\textbf{k}}+\mu\end{pmatrix},
	\end{align}
	where $\Delta_{\textbf{k}}=\dfrac{\Delta_0}{2} (\cos(k_{x})-\cos(k_{y}))$ is the pairing gap, $\varepsilon_{\textbf{k}}$ is the band structure of the material, and $\mu$ the chemical potential \cite{Note1}.

	We now introduce a minimal model for CDW modulations, by considering the scattering from a momentum-independent charge impurity, $\rm V_\textbf{k} = V_0\sigma^z$. In this case, the integral over $\omega$ in Eq.~(\ref{eq:rhoq}) delivers
	\begin{align} 
	\chi(\textbf{q},\Omega) =&~ 2\pi V_{0}\int d^2 k~ \dfrac{E_{\textbf{k}}^2+\varepsilon_{\textbf{k}}\varepsilon_{\textbf{k}+\textbf{q}}-\Delta_{\textbf{k}}\Delta_{\textbf{k}+\textbf{q}}-E_\textbf{k}\Omega}%
	{E_{\textbf{k}}[E_{\textbf{k}+\textbf{q}}^2-(E_{\textbf{k}}-\Omega)^2]} \nonumber\\%
	 & +\dfrac{E_{\textbf{k}+\textbf{q}}^2+\varepsilon_{\textbf{k}}\varepsilon_{\textbf{k}+\textbf{q}}-\Delta_{\textbf{k}}\Delta_{\textbf{k}+\textbf{q}}+E_{\textbf{k}+\textbf{q}}\Omega}{E_{\textbf{k}+\textbf{q}}[E_{\textbf{k}}^2-(E_{\textbf{k}+\textbf{q}}+\Omega)^2]},\label{eq:rhoq2}
	\end{align} 
	where $E_{\textbf{k}} = \sqrt{\varepsilon_{\textbf{k}}^2+\Delta_{\textbf{k}}^2}$. 
	In the limit of $\Delta_\textbf{k}\to0$, one has $E_\textbf{k}=|\varepsilon_\textbf{k}|$ and Eq.~(\ref{eq:rhoq2}) recovers the Lindhard response function of free fermions used in Ref.~\cite{dalla2016friedel}.
	
	To describe X-ray scattering experiments of cuprates, we use a tight-binding model, $\varepsilon_{\textbf{k}}=-2t[\cos(k_{x})+\cos(k_{y})]-4t'\cos(k_{x})\cos(k_{y})$, where $t$ and $t'$ are nearest neighbor (NN) and next-nearest neighbor (NNN) hopping coefficients. The parameter $t'$ strongly affects the shape of the Fermi surface: superconducting cuprates are close to half-filling and, for $t'=0$, their Fermi surface has a diamond shape. A negative $t'$ leads to a Fermi surface with parallel segments (nesting) at the antinodal wavevectors ${\bf  k} = (\pm\pi/a,0)$ and $(0,\pm\pi/a)$. As pointed out long ago \cite{massidda1989electronic}, these parallel segments are prone to induce finite wavevector instabilities, such as CDWs and PDWs. This approach matches the experimentally observed doping dependence of the wavevector (see the Supplemental Materials)\cite{Note2}.

	Let us first consider the elastic component ($\Omega=0$), by comparing Eq.~(\ref{eq:rhoq2}) with resonant  X-ray scattering experiments of three different cuprates: electron-doped NCCO \cite{jang2017superconductivity}
	, underdoped BSCCO \cite{comin2014charge}
	and underdoped Hg1201 \cite{tabis2014charge}.
	The corresponding plots are shown as blue curves in Fig.~\ref{Fig1}, where we select $\mu$ to match the experimental doping $x$.  The superconducting gap $\Delta_0$ has a minor influence on these plots and is set to physically relevant values. The fitting parameter $t'$ is obtained by minimizing the difference between the theoretical curves and the actual experiments \cite{Note3}. The values of $t'/t$ obtained by this procedure are consistent with the Fermi surfaces determined by ARPES \cite{norman1995phenomenological, markiewicz2005one}. For all three materials, we obtain an excellent agreement between the theoretical curves and the experiments: Eq.~(\ref{eq:rhoq2}) describes well both the period of the modulation and the width of the peak. 
	
	As mentioned above, an alternative explanation for the observed signal are PDW fluctuations. Specifically, we consider short-ranged PDWs that coexist with a static and uniform (d-wave) pairing gap $\Delta_0$. Our analysis does not apply to materials where the PDWs are long ranged and give rise to a state, analogous to the FFLO state \cite{fulde1964superconductivity, larkin1965nonuniform}, where the pairing gap is periodically modulated in space  (such as the striped superconductor LBCO near 1/8 doping \cite{berg2009theory, berg2009striped}). Following our weak-coupling approach, we consider PDW fluctuations induced by
	%
	a local modulation of the pairing gap, $\rm {V_\textbf{k}} = \Delta_\textbf{k}\sigma^{x}$ in Eq.~(\ref{eq:rhoq}), where $\rm \sigma^{x}$ is a Pauli matrix.
	By performing the integral over $\omega$ in Eq.~(\ref{eq:rhoq}), we obtain
	\begin{align}\label{eq:rhoq3}
	\chi(\textbf{q},\Omega) =2\pi \int d^2 k~ \Delta_{\bf k}\frac{\varepsilon_{\bf k}\Delta_{\bf k+q}+\varepsilon_{\bf k+q}\Delta_{\bf k}}{(E_{\bf k}-E_{\bf k+q})^2-\Omega^2}\left(\dfrac1{E_\textbf{k}}-\dfrac1{E_{\textbf{k}+\textbf{q}}}\right).
	\end{align}
	The resulting plots are shown as red curves in Fig.~\ref{Fig1}. We find that the PDW signal shows pronounced peaks at approximately the same wavevector as the CDW one. The precise shape of the peaks depends on the details of the band structure and cannot be used to identify the type of modulation. As a result, one-dimensional scans of the X-ray scattering are not sufficient to distinguish unequivocally between CDW and PDW fluctuations.

	\begin{figure} [t]
		\centering
		\begin{tabular}{c c}
			(a) CDW & (b) PDW \\
			\includegraphics[width=0.25\textwidth]{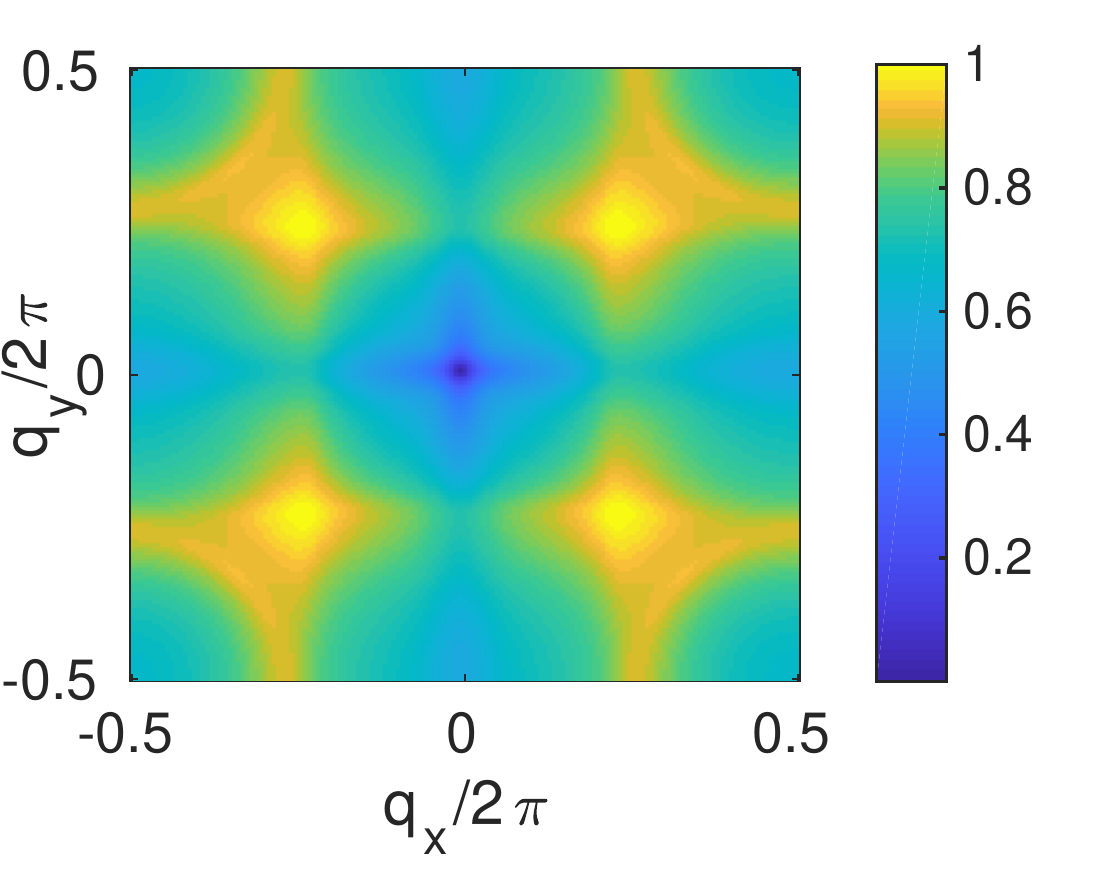} &
			
			\includegraphics[width=0.25\textwidth]{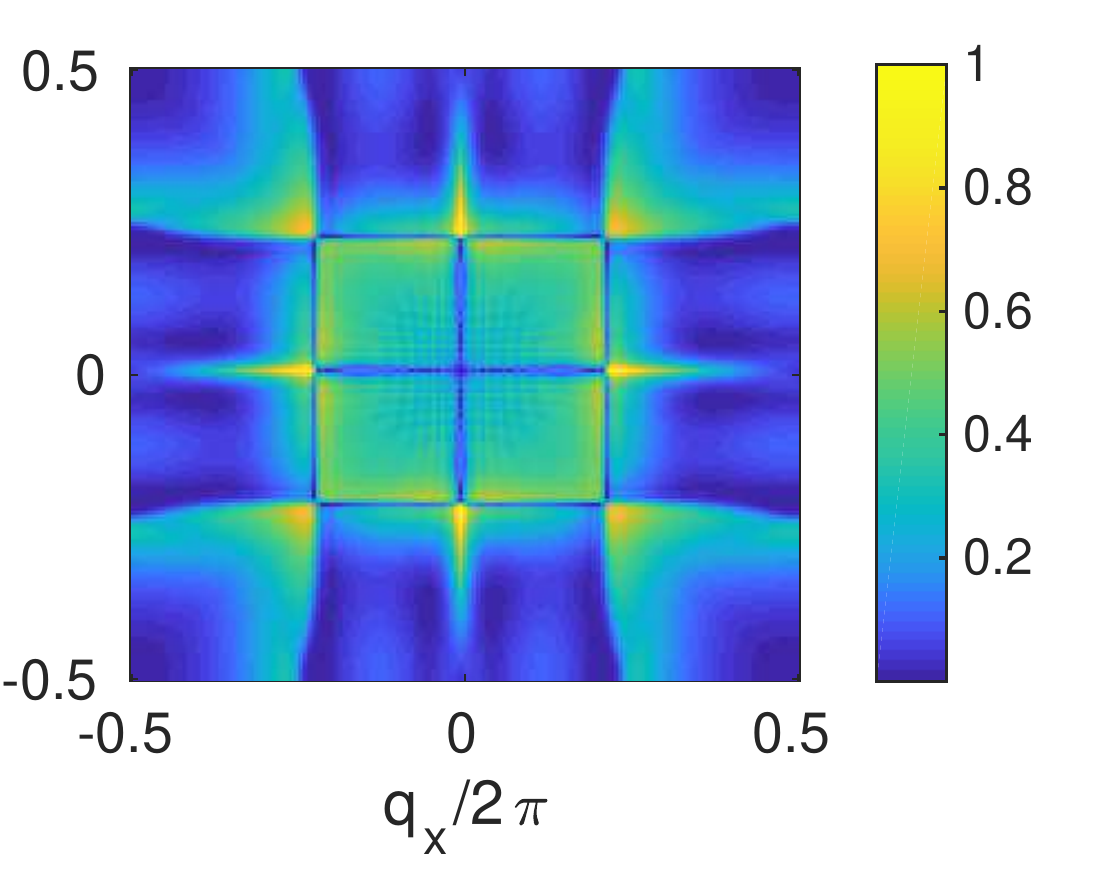}	

		\end{tabular}
		\caption{Two dimensional maps of the elastic response $\chi(\textbf{q},\Omega=0)$ in the weak-coupling approach: (a) CDW, Eq.~(\ref{eq:rhoq2}), (b) PDW, Eq.~(\ref{eq:rhoq3}). Numerical parameters: $ x=0.14, ~t'/t=-0.6,~ \Delta_{0}/t = 0.2$.  }
		\label{fig:CPDW}
	\end{figure}

	{\it Identifying CDW and PDW --} We now present two distinct methods to distinguish between these two types of modulations, based respectively on the momentum and energy dependence of the X-ray scattering signal. The first method uses the full two dimensional map of $\chi(\textbf{q},\Omega=0)$. Two representative theoretical maps are shown in Fig.~\ref{fig:CPDW}. Although both maps have pronounced peaks at the same wave-vector ($q \approx \pm0.25$), their two-dimensional structure is very different: The CDW signal has four peaks at $\textbf{q} = (\pm q,\pm q)$ and four saddle points at $\textbf{q} = (0,\pm q)$ and $(\pm q,0)$. In contrast, the PDW signal has four strong peaks at $\textbf{q} = (0,\pm q)$ and $(\pm q,0)$ and four weaker peaks at $\textbf{q} = (\pm q,\pm q)$. We claim that the ratio between the intensity of the signal at these two wavevectors can be used to identify the type of modulation. For the parameters used in Fig.~\ref{fig:CPDW}, we find
	that $R\equiv \chi(\textbf{q}=(q, 0), \Omega=0)/\chi(\textbf{q}=(q, q), \Omega=0) \approx 0.7$ for CDW and $R\approx 1.4$ for PDW.
	This result is very robust: Although the precise value of $R$ depends on the microscopic parameters of the model, we find generically that $R<1$ for CDWs and $R>1$ for PDWs.

	To understand this result, we recall that in our weak-coupling model of CDW and PDW, the scatterers are local and isotropic and, hence, their scattering matrices are momentum independent. In this minimal model, the intensity of the response function only depends on the density of states, i.e.~on the shape of the Fermi surface and on the symmetry of the pairing gap. As mentioned before, the Fermi surface of cuprates has 4 pairs of parallel segments, leading to a doping-dependent nesting wavevector $q$ . Interestingly, the nesting at wavevector $(q,q)$ is more effective than at wavevector $(q,0)$: In the former case one obtains an overlap between all 4 pairs of segments of the Fermi surface, while  in the latter only 2 pairs are involved. This observation explains why the CDW is more pronounced at wavevector $(q,q)$ than at $(q,0)$,  i.e. $R<1$ \cite{Note4}.  In the case of the PDW signal, Eq.~(\ref{eq:rhoq3}), each segment of the Fermi surface is weighted by the corresponding value of $\Delta_\textbf{k}$. This factor strongly favors the wavevector $(q,0)$, which connects antinodes to antinodes, with respect to $(q,q)$, which connects antinodes to nodes. Hence, for PDWs $R>1$ in agreement with the numerical result mentioned above.

	\begin{figure}[t]
			\begin{tabular}{c c}
		(a) CDW & (b) PDW \\
		\includegraphics[width=0.25\textwidth]{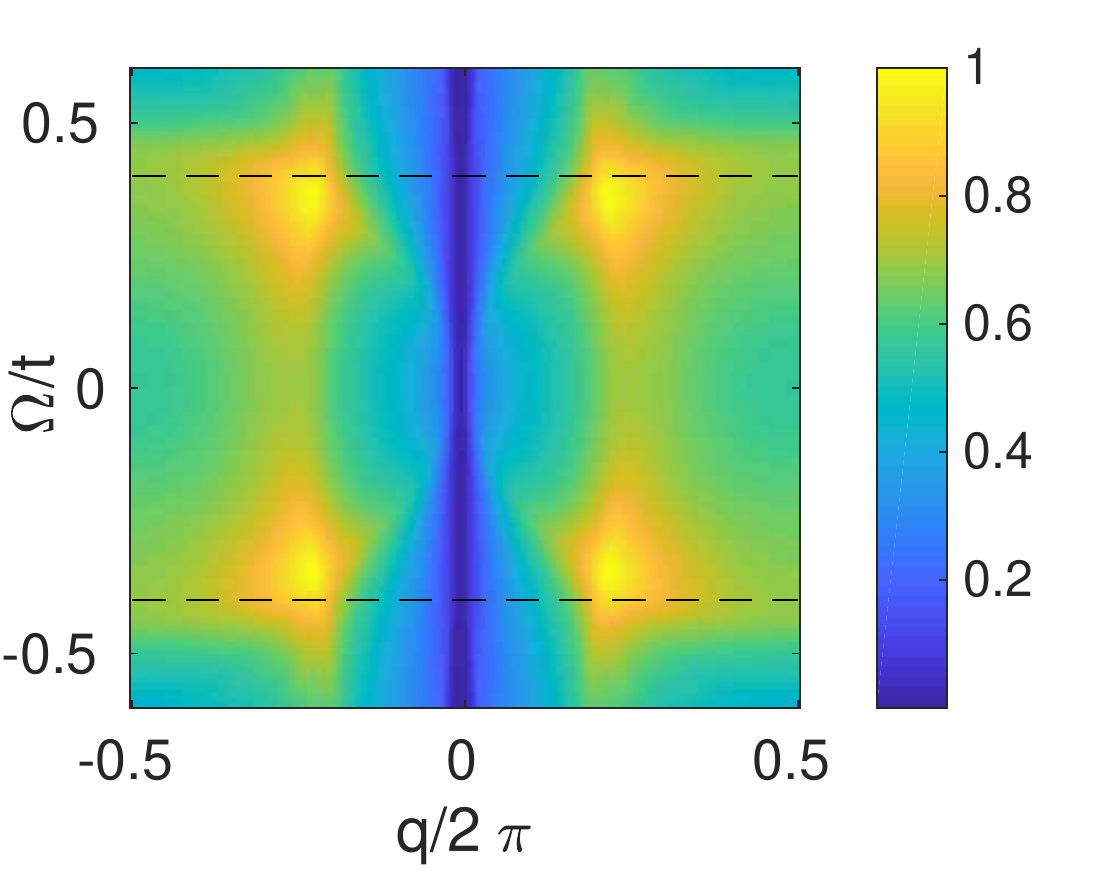} &
		\includegraphics[width=0.25\textwidth]{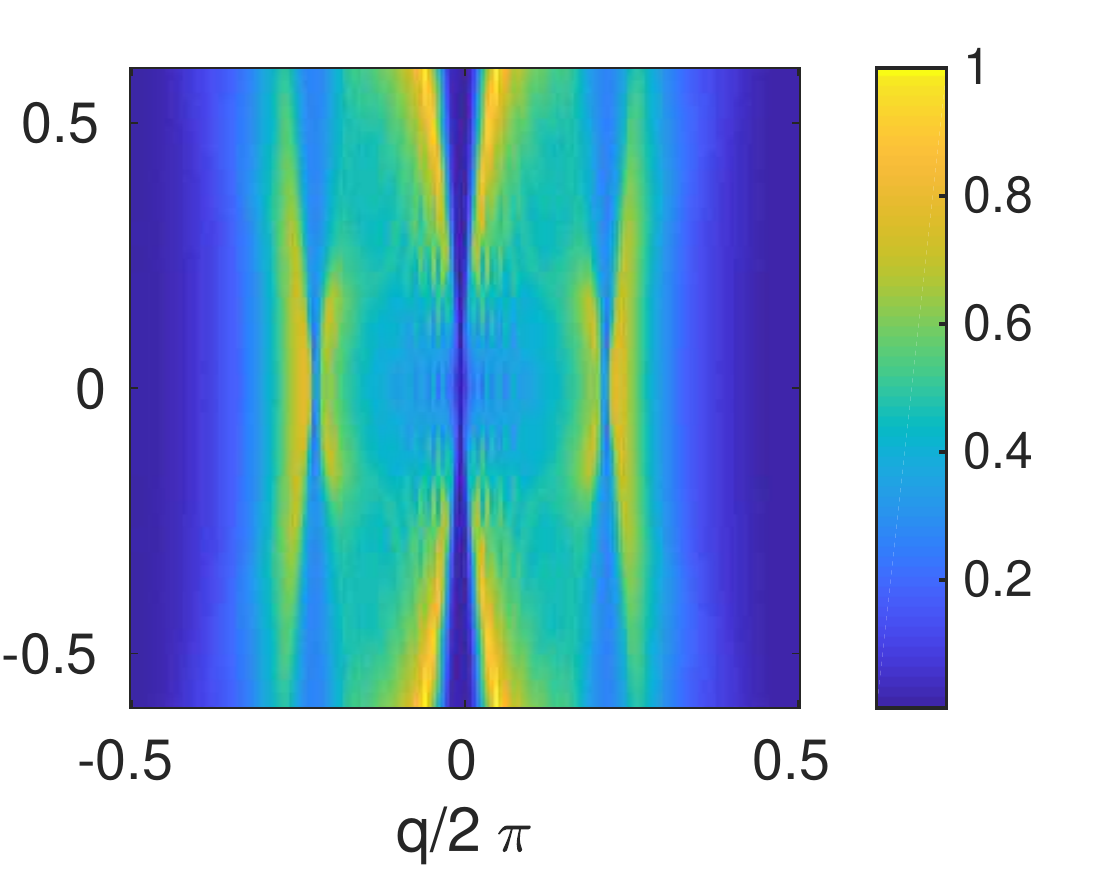} 
		\end{tabular}
		\caption{Energy-momentum dependence of the inelastic response $\chi(\textbf{q},\Omega)$ along the line $(q, 0)$ in the weak coupling approach for (a) CDWs, Eq.~(\ref{eq:rhoq2}), and (b) PDWs, Eq.~(\ref{eq:rhoq3}). Same numerical values as in Fig.~\ref{fig:CPDW}. The black dashed lines in (a) are located at $\Omega/t = \pm 2\Delta_{0}/t$.} 
		\label{Fig:inelastic} 
	\end{figure}

	The experimental data strongly supports the PDW scenario: (i) Transverse and longitudinal one-dimensional scans in the vicinity of $(q,0)$ show that the scattering amplitude peaks in both directions \cite{peng2018re}. This experimental finding is in stark contradiction with the CDW case, where a saddle-point is expected, and agrees with the  PDW case (see Fig.~\ref{fig:CPDW}). (ii) The peak at $(q,q)$ is small \cite{GhiringhelliPrivate} or absent \cite{arpaia2019dynamical, kang2019evolution}, indicating that $R>1$, whereas for CDW this should be the dominant peak. Both observations are consistent with the PDW scenario only. 
	
    {\it Energy dependence -- }Let us now turn to the energy dependence of the response function $\chi(\textbf{q},\Omega)$. Fig.~\ref{Fig:inelastic} shows our theoretical predictions for incipient CDWs, Eq.~(\ref{eq:rhoq2}), and PDWs, Eq.~(\ref{eq:rhoq3}). The energy dependence of the two graphs is very different: $\chi(\textbf{q},\Omega)$ is peaked at $\Omega \approx  2\Delta_{0}$ for CDWs and at $\Omega=0$ for PDWs. This discrepancy can be rationalized by noting that charge impurities create particle-hole pairs and, hence, need to overcome the energy gap $\Delta$. In contrast, local modulations of the pairing gap can create two holes (or two particles) at the same energy, below (or above) the gap. As a consequence, the response to CDWs is peaked at $2\Delta_0$, while the response to PDWs is peaked at zero energy. Recent energy-resolved inelastic X-ray scattering (RIXS) experiments \cite{ghiringhelli2012long,chaix2017dispersive,peng2018re,da2018coupling,arpaia2019dynamical} show that the signal is peaked at (or close to) zero energy and, again, favor the PDW scenario. Furthermore, our theoretical model accounts for the experimental observation of a dispersive peak that departs from the zero-energy peak towards higher energies~\cite{chaix2017dispersive}.

	{\it Hubbard model -- }The weak coupling approach considered above does not take into account the interactions between quasiparticles, which can enhance the CDW and PDW fluctuations and lead to a competition between them. To capture these effects,  we now consider a two-dimensional extended Hubbard model with on-site repulsion $U$ and NN attraction $V$. This model shows several competing phases, such as the Mott insulator and the d-wave superconductor, that are generic to cuprates  \cite{micnas1988extended, micnas1990superconductivity,monthoux1994self,newns1995van, husslein1996quantum, takigawa2004quasiparticle}. Under the usual mean-field approximation $n_{j}=\sum_{\sigma=\pm}\langle c^{\dagger}_{j, \sigma}c_{j, \sigma} \rangle/2$,  $\Delta_{\hat{e}, j}=\sum_{\sigma=\pm}\sigma\langle c_{j, -\sigma}c_{j+\hat{e}, \sigma} \rangle/2$, 
	the Hamiltonian reads
	\begin{align}\label{eq:hubbard}
	H&=-t\sum_{\langle i, j \rangle,\sigma} c^{\dagger}_{i, \sigma} c^{\phantom{\dagger}}_{j, \sigma} -t'\sum_{\langle\langle i, j \rangle \rangle,\sigma} \left(c^{\dagger}_{i, \sigma} c^{\phantom{\dagger}}_{j, \sigma} 
	+\mathrm{H.c.}\right)\\ \nonumber &+ U \sum_{j, \sigma} n^{\phantom{\dagger}}_{j} c^{\dagger}_{j, \sigma} c^{\phantom{\dagger}}_{j, \sigma}  + V \sum_{\hat{e}, j,\sigma}\left( \Delta^{\phantom{\dagger}}_{\hat{e}, j} c^{\dagger}_{j, \uparrow} c^{\dagger}_{j+\hat{e}, \downarrow} +\mathrm{H.c.} \right) 
	\end{align}
	where $\langle\cdot,\cdot\rangle$ denotes NN, $\langle\langle\cdot,\cdot\rangle\rangle$ denotes NNN, $\hat{e}$ connects NN sites, and H.c.~stands for Hermitian conjugate. For $V<0$, the self-consistent solution of Eq.~(\ref{eq:hubbard}) delivers a superconductor with d-wave order parameter $\Delta_j = \frac{1}{4} (\Delta_{\hat{x}, j}+\Delta_{-\hat{x}, j} - \Delta_{\hat{y}, j}-\Delta_{-\hat{y}, j})$.

	\begin{figure} [t]
		\centering
		
		\begin{tabular}{c c}
			(a) site impurity ($\delta U$) & (b) bond impurity ($\delta V$) \\
			\includegraphics[width=0.25\textwidth]{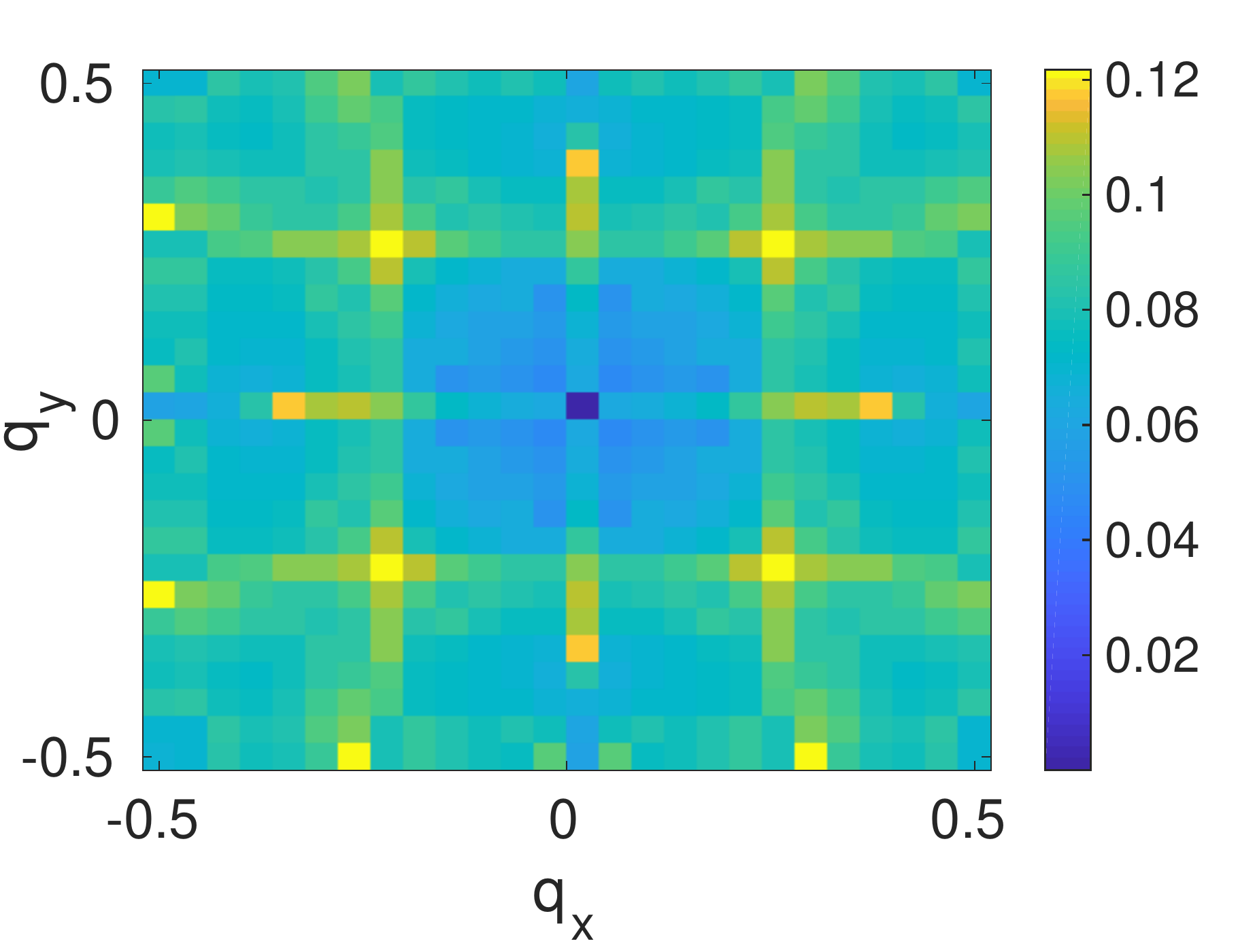} &
			\includegraphics[width=0.25\textwidth]{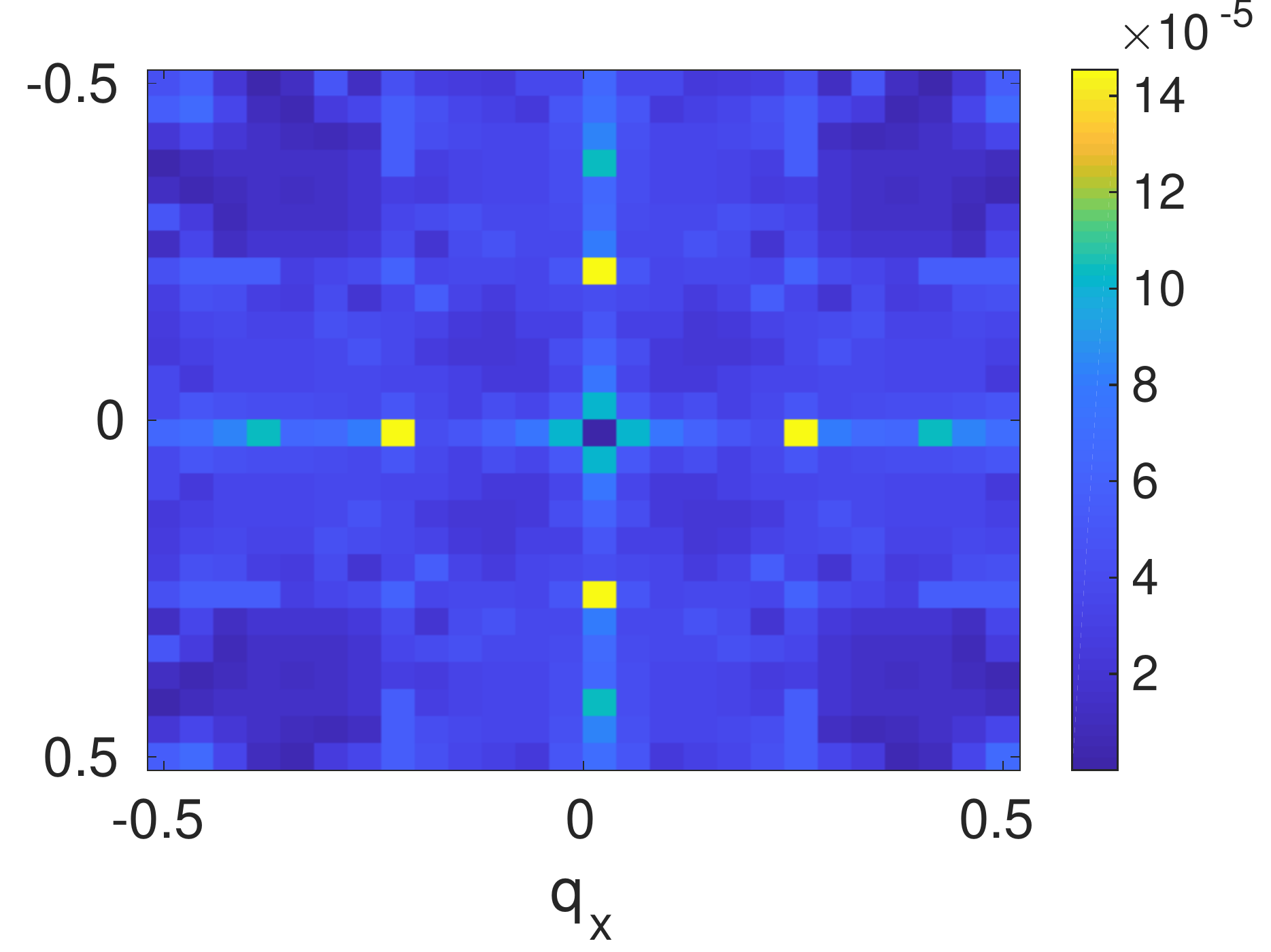}
		\end{tabular}
		\caption{Two dimensional maps of the density fluctuations $n_\textbf{q}$ in the Hubbard model with a local modulation of the interaction: (a) on a single site, (b) on a single bond. For clarity, subfigure (b) has been symmetrized by 90 degrees. Numerical parameters: $x=0.16$, $t'/t=-0.6$ and $U/t=1.5, V/t=-0.5, ~\delta U/t = \delta V/t = 0.2$.}	
		\label{fig:hubbard}
	\end{figure}

	 To study the interplay between CDW and PDW we add a local impurity, which generically leads to spatial modulations of both the charge $n_{j}$ and the pairing gap $\Delta_{j}$. We classify the resulting density wave as CDW or PDW depending on which modulation is dominant, by comparing the relative standard deviations ${\delta n}$ and ${\delta \Delta}$ \cite{Note5}. For simplicity, we focus here on two representative impurities (see the Supplemental Materials for additional examples): a single site with $U \to U+\delta U$ and a single bond with $V \to V+\delta V$. In the former case, we find that the density wave has mixed CDW/PDW character ($\delta\Delta/\delta n \approx 1 $), while in the latter case it has a predominant PDW character ($\delta \Delta/\delta n \approx 140$).
	
	To model the response to X-ray scattering, we now focus on the  Fourier transformed density $n_{\bf q}$, Fig.~\ref{fig:hubbard}. As expected, we find that interactions enhance density wave instabilities and lead to narrower peaks, with longer range correlations. In our calculations, the width of these peaks is limited by the system size ($L=26$ unit cells), suggesting that the Hubbard model is consistent with long ranged CDW/PDW modulations. By comparing the two subplots, we observe that the on-site impurity leads to peaks at both wavevectors $(q,q)$ and $(q,0)$, while the bond impurity leads to pronounced peaks in at the wavevector $(q,0)$ only. This is consistent with our proposal to compare the intensities of the two peaks to distinguish between CDW and PDW modulations.

 {\it Discussion -- }In this paper we described a method to distinguish between incipient CDW and PDW fluctuations, based on the analysis of X-ray scattering experiments. When these spatial modulations coexist with a homogeneous superconducting gap $\Delta_0$, both oscillations couple directly to any physical observable, such as the charge density and the tunneling density of states. Hence, both CDWs and PDWs can be detected using different experimental technique, including X-ray scattering, scanning tunneling spectroscopy (STS), and scanning Josephson probes. Here we focused on X-ray scattering and showed that one-dimensional cuts are not unequivocal, because they can be adequately fitted by both types of modulations. In contrast, two-dimensional maps of CDWs and PDWs are very different: the former are peaked at wavevector $(q,q)$, while the latter has stronger peaks at $(q,0)$. These two types of density waves can be further distinguished by energy-resolved RIXS measurements: CDWs are peaked at $\Omega=2\Delta_ {0}$, while PDWs are peaked at $\Omega=0$.

 These results strongly rely on our simplifying assumption of noninteracting quasiparticles, scattered by local and isotropic pinning centers. In principle, other sources of inhomogeneity, as well as strong electronic correlations, can yield different results and enhance CDW and PDW signals in other directions. To address this point, we considered an attractive Hubbard model, which demonstrated that our method to identify fluctuations with a dominant CDW or PDW character remains valid in the presence of strong interactions. In particular, we showed that local modulations of the pairing mechanism (in our case, a nearest-neighbor attractive interaction) give rise to density waves peaked at $(q,0)$ and with a predominant PDW character.
	
	Our theoretical model of PDW reproduces the main features of recent X-ray scattering experiments of superconducting NCCO, Hg1201, and BSCCO. Specifically, our theory explains why (i) the X-ray scattering signal is peaked at wavevector $(q,0)$, rather than at $(q,q)$, as expected for CDWs; (ii) the RIXS signal is peaked at frequency $\Omega=0$ and is accompanied by weaker dispersive inelastic peaks.  Our findings also agree with earlier STS experiments of BSCCO \cite{howald2003periodic, vershinin2004local, hanaguri2004checkerboard}, which found that the incommensurate checkerboard order has a dominant PDW character \cite{pereg2003theory,nowadnick2012quasiparticle,dalla2015exploring,dai2018pair}.  Attributing the X-ray signal to fluctuations of the pairing order parameter (PDW) explains its temperature dependence: these fluctuations are strongest at the critical temperature of superconducting order parameter, $T_c$, in agreement with the experimental observations (\cite{chang2012direct,da2014ubiquitous}).
	 Finally, The proposed PDW scenario explains why the signal detected in X-ray scattering of cuprates is orders of magnitude smaller than the one observed in ordinary CDW materials ($\delta n\ll \delta \Delta$). 
	
	Our method can be further extended to include the effects of magnetic fields by considering a Hubbard model with complex hopping elements (Peierls substitution). In type-II superconductors, external magnetic field generate isolated vortices in whose core the pairing gap is locally suppressed. Hence, a magnetic vortex acts as a pinning site for a PDW modulation, in analogy to the bond impurity considered in this paper. Numerical studies of the Hubbard model in the presence of magnetic fields have indeed found that spatial modulations of the pairing gap develop in the proximity of the vortex core \cite{zhu2002spin,takigawa2004quasiparticle,simonucci2013temperature}. This finding is consistent with STS experiments demonstrating that the periodic modulations are mostly pronounced in the vicinity of the vortex core \cite{hoffman2002four, matsuba2007anti, yoshizawa2013high,edkins2019magnetic}, as well as with evidence that the density waves become long ranged at high magnetic fields \cite{doiron2007quantum,wu2011magnetic,wu2013emergence,gerber2015three}. By locally suppressing superconductivity, large densities of magnetic vortices can lead to a long ranged PDW order.

	\begin{acknowledgments}
		We thank Peter Abbamonte, Lucio Braicovich, Debanjan Chowdhury, Riccardo Comin, J.~C.~Seamus Davis, Andrea Damascelli, Eugene Demler, Giacomo Ghiringhelli, Marco Grilli, Jenny Hoffman, Amit Keren, Steve Kivelson, and Dror Orgad  for useful discussions. This work is supported by the Israel Science Foundation grants No. 151/19 and 967/19.
	\end{acknowledgments}

\end{document}